\documentclass[sigconf]{acmart} 
\AtBeginDocument{%
  }

\begin{document}

\title{Benchmarking Traditional and Learning-Based Reference-Free Compressors for Long-Read Sequencing Data}


\author{Huidong Ma}
\authornote{Both authors contributed equally to this research.}
\email{mahd@nbjl.nankai.edu.cn}
\affiliation{%
    \institution{Nankai University}
    \department{College of C.S., Nankai-Baidu Joint Lab, TMCC, SysNet, DISSec, GTIISC}
    \city{Tianjin}
    \country{China}
}

\author{Hui Sun}
\authornotemark[1]
\affiliation{%
    \institution{Nankai University}
    \department{College of C.S., Nankai-Baidu Joint Lab, TMCC, SysNet, DISSec, GTIISC}
    \city{Tianjin}
    \country{China}
}

\author{Xinyan Shi}
\affiliation{%
    \institution{Nankai University}
    \department{College of C.S., Nankai-Baidu Joint Lab, TMCC, SysNet, DISSec, GTIISC}
    \city{Tianjin}
    \country{China}
}

\author{Cheng Zhong}
\affiliation{%
    \institution{Guangxi University}
    \department{School of Computer, Electronics and Information}
    \city{Nanning}
    \country{China}
}

\author{Xiaoguang Liu}
\authornote{Corresponding authors}
\affiliation{%
    \institution{Nankai University}
    \department{College of C.S., Nankai-Baidu Joint Lab, TMCC, SysNet, DISSec, GTIISC}
    \city{Tianjin}
    \country{China}
}

\author{Gang Wang}
\authornotemark[2]
\affiliation{%
    \institution{Nankai University}
    \department{College of C.S., Nankai-Baidu Joint Lab, TMCC, SysNet, DISSec, GTIISC}
    \city{Tianjin}
    \country{China}
}

\renewcommand{\shortauthors}{Huidong Ma et al.}


\begin{abstract}
The development of Long-Read Genomics Sequencing (LRGS) technologies has led to a significant increase in biological sequencing data.
Compressing LRGS is an important measure for large-scale data sharing, transmission, and storage. Although some reference-based compressors have been proposed to save storage space for LRGS data, the use of additional reference genome limits their widespread adoption, making reference-free methods still the mainstream.
However, selecting a suitable reference-free compressor for complex application scenarios, such as real-time transmission and long-term backup of LRGS data, is challenging due to the lack of comprehensive reviews and systematic benchmarking of these methods.
To this end, 1) we first introduce state-of-the-art general-purpose and LRGS-dedicated compression methods, especially emerging learning-based compressors; 2) we perform a comprehensive benchmark evaluation on 35 classical algorithms using 16 real-world datasets with differing sequencing platforms, species, and lengths; 3) we analyze the overall performance of each compressor by several metrics, including compression ratio, running time, memory overhead, robustness performance, et al. In addition, 4) for future LRGS data compressors, we outline investigation directions with consideration for privacy-sensitive sequence data security, hardware parallel acceleration, parameter tuning framework, and system hardware-algorithm integration design. We organize all evaluation results into LRCBench (\underline{L}ong-\underline{R}ead \underline{C}ompression \underline{Bench}mark), which will be continuously updated and maintained. The source-code and datasets are available at \url{https://github.com/huidong-ma/LRCBench}.
\end{abstract}



\maketitle

\section{Introduction}
The widespread adoption of high-throughput and third-generation sequencing technologies has significantly reduced the expenditure of acquiring big genomics sequencing data~\cite{sra,cost}. 
In 2002, sequencing cost around \$5,292.39 per megabytes (MB), but by 2022, it decreased to \$0.006 per MB~\cite{cost}. Consequently, there has been a significant increase in bio-data globally due to the rapidly declining expenses of sequencing. 
For example, as of February 2025, the 17,556 terabytes (TB) of sequencing data had been stored in the China National Gene Bank Sequence Archiving System (CNSA)~\cite{cnsa}, and as of June 2022, the National Center for Biotechnology Information Sequence Read Archive (SRA) has backed up more than 20,760 TB of sequencing data~\cite{ncbi-sra}. 
Among big bio-data, long-read sequencing technologies, such as Pacific Biosciences (PacBio) and Oxford Nanopore Technologies (ONT), can generate read lengths more extraordinary than 10 kbp, which is more extended than short-read generated by high-throughput sequencing approaches~\cite{PacBio,ONT,HU2021801}. 
These Long-Read Genomics sequencing (LRGS) data generation technologies can overcome short reads issues, such as structural variant detection and genome-wide repeats~\cite{HU2021801,sedlazeck2018accurate,heller2019svim,jiang2020long,ma2023cnnlsv,ma2023ricme}. 

Third-generation sequencing technologies hold higher throughput and longer sequencing lengths that have increased the use of long-read and the scale of generated data~\cite{goodwin2016coming,zhang2020}. Developing LRGS-dedicated lossless compressors is critical in reducing data storage requirements and lowering the infrastructure storage expenditure. However, due to the relatively short period since the emergence of long-read sequencing technologies, academia and industry have focused on the compression of high-throughput short reads, assembled genomics, and quality score data, over the past three decades~\cite{jizheng2010dna,zhu2015high,numanagic2016comparison,hernaez2019genomic,kredens2020vertical,kryukov2020sequence,gilmary2021compression,sun2024lrcb}. 
The traditional fixed-length short-read compressors hold severe limitations in compressing variable-length long reads with high sequencing error rates. Recently, computational biologists have been emphasizing the design of algorithms for compressing long-read data~\cite{wang2023csnmg,kokot2022colord,renano,meng2023reference}. The LRGS-dedicated compressors include reference-based and reference-free methods. The reference-based methods achieve the most satisfactory compression effects in reducing the size of the original data at the cost of additional calculations and optimal selection of the reference genome~\cite{kokot2022colord,renano}. However, the unavailability of an appropriate reference genome frequently limits the utilization of reference-based compressors. As a result, research endeavors in this domain predominantly revolve around reference-free long-read data compressors.

Previous investigations concentrated on benchmarking short-read data, FastA and FastQ format file compression tools~\cite{kryukov2020sequence,numanagic2016comparison,kredens2020vertical}, lacking professional evaluation for LRGS data compression (LRCB~\cite{sun2024lrcb} is the only benchmark that focuses on LRGS data compression, but it lacks comprehensive reviews and testing). 
In addition, learning-based compressors have emerged in the field of data compression~\cite{MSDLC,MSDZip,ma2026efficient,ma2026learned}. The advent of large language model such as DeepSeek Series~\cite{lu2024deepseekv1,liu2024deepseekv2,liu2024deepseekv3}, ChatGPT~\cite{achiam2023gpt}, and LLaMA Series~\cite{touvron2023llama,touvron2023llama2,dubey2024llama} has brought new opportunities and challenges to LRGS compression.
Therefore, it is necessary for comprehensive benchmark evaluation for lossless reference-free LRGS data compressors, specifically for those emergence learning-based methods~\cite{NNLCB}. To stimulate the development of more efficient sequencing big data compression tools, we propose the LRCBench (\underline{L}ong \underline{R}eads \underline{C}ompression \underline{Bench}mark), which is a comprehensive evaluation of the most state-of-the-art reference-free lossless compressors for LRGS data. The main innovations and contributions of this paper are summarized as follows.
\begin{itemize} 

\item We systematically review the relevant work on reference-free compression of LRGS data, including traditional general-purpose methods and LRGS-dedicated methods, especially emerging learning-based compressors. 

\item We propose LRCBench, a comprehensive benchmarking 35 compressors on 16 datasets of different types and sizes; the evaluation metrics include compression ratio, running time, memory consumption, robustness performance, et al.

\item We summarize potential development directions for LRGS compression, including: 1) parallel acceleration of compressors, 2) privacy-sensitive security compression for large-scale LRGS data, 3) self-adaptive parameter optimization frameworks, and 4) hardware-algorithm integration design.
\end{itemize}

\section{Long-Read Genomics Sequencing (LRGS) Data Compression Methods}~\label{S2}
In terms of the types of data supported by compressors, we categorize LRGS compressors into two groups: general-purpose and LRGS-dedicated. However, with the development of deep learning technology, several learning-based general-purpose and dedicated compressors have emerged that combine neural networks with entropy coding algorithms to achieve better lossless compression. Therefore, in this paper, we further subdivide these compressors into four categories:
1) traditional general-purpose compressors;
2) learning-based general-purpose compressors;
3) traditional LRGS-dedicated compressors;
4) learning-based LRGS-dedicated methods.

\subsection{\textbf{Traditional General-purpose Compressor}}
Many of the current Traditional General-purpose Compressors (TGCs) are improvements over the basic classical compression methods (LZ77~\cite{ziv1977universal}, LZ78~\cite{ziv1978compression} and BWT~\cite{1994BWT} et al.) and entropy coding algorithms (Huffman Coding~\cite{huffman-encoding}, Arithmetic Coding~\cite{witten1987arithmetic} and Asymmetric Numeral Systems~\cite{duda2013asymmetric}, et al). The basic compression methods reduce data redundancy by finding patterns of repeated occurrences in the data. Taking LZ77 as an example, the basic idea of the algorithm is to reduce the amount of data by finding strings that recur in the data and replacing these strings with pointers to previously occurring locations. Entropy coding algorithms~\cite{huffman-encoding,witten1987arithmetic} can further improve the compression rate on the basis of the basic compression algorithm by analyzing the frequency of occurrence of different symbols or patterns in the data and using different lengths of codes to represent symbols of different frequencies.

Brieflz~\cite{Brieflz}, Brotli~\cite{Brotli}, GZip~\cite{Gzip}, Lizard~\cite{Lizard}, LZ4~\cite{LZ4}, LZMA~\cite{LZMA}, LZO~\cite{LZO} and ZSTD~\cite{zstd} are all general-purpose compression algorithms that improve on the LZ77 algorithm. Among them, Brotli, GZip, Lizard and ZSTD are all based on LZ77 and further combined with Huffman Coding to achieve better compression ratio. PigZ is a parallelized version of GZip, which uses the same DEFLATE~\cite{deutsch1996rfc1951} algorithm (LZ77+Huffman Coding) as GZip, but splits a large file into multiple blocks through multi-threading. LZOP~\cite{lzop} is based on LZO, using the LZ77 algorithm and fast dictionary lookup to achieve compression. Compared to LZO, LZOP further optimizes speed and dictionary management by adjusting the details of the algorithm. LZMA introduces Range Coding (RC)~\cite{lelewer1987data} as an alternative to the commonly used Huffman coding to achieve higher compression ratios, but at the same time has slower compression speeds. LZMA2~\cite{LZMA,fast-lzma2} is a multi-threaded version of LZMA that uses the same compression method as LZMA, but introduces support for multi-core processors to parallelize and speed up the compression process. SnZip~\cite{Snzip} is based on the Snappy~\cite{Snappy} algorithm and focuses on very high decompression speeds. Snappy uses a similar technique to LZ77, but it focuses on increasing decompression speeds by reducing the number of computations used in the compression process.

Both BZip2~\cite{Bzip2} and its variant BSC~\cite{BSC} use a combination of Burrows-Wheeler Transform (BWT)~\cite{1994BWT} and Move-To-Front (MTF)~\cite{ryabko1980data} algorithms combined with Huffman coding to achieve further compression.
BWT focuses on clustering character repetitions by reordering characters, and then further reduces data redundancy through MTF and Huffman coding to achieve efficient compression. PBZip2~\cite{PBZIP2} is a parallel implementation of BZip2, which still uses BWT and MTF transform in the compression process, but it greatly improves the compression speed by splitting the data into multiple parts through multi-threading.

PPMD~\cite{PPMD} is a statistics-based compressor that utilizes the Prediction by Partial Matching (PPM). It encodes text data more efficiently by constructing a prediction model that predicts the probability of occurrence of the current character based on the historical frequency of the previous characters.
ZPAQ~\cite{ZPAQ} uses an incremental method based on dictionary and block reuse. It is characterized by its support for incremental backups, i.e., it compresses only the new or modified data without recompressing all the data.

\subsection{\textbf{Learning-based General-purpose Compressors}}
Similar to the statistics-based methods PPMD~\cite{PPMD} and ZPAQ~\cite{ZPAQ}, Learning-based Lossless Compressors (LGCs) also contain two core parts: probability prediction model and entropy encoder. 

In the probability prediction phase, the neural network model predicts the probability distribution of the target symbol based on the contextual relationship of history symbols. Unlike traditional compressors, existing learning-based compressors usually use arithmetic coding~\cite{witten1987arithmetic} as the entropy encoder in pursuit of better compression ratios. Since arithmetic coding algorithm has converged to the entropy limit of lossless compression, existing work usually improves the probability prediction model. More accurate prediction model will lead to lower entropy information for better compression. Currently, common probability prediction models include Recurrent Neural Networks (RNN)~\cite{zaremba2014recurrent}, attention mechanisms~\cite{vaswani2017attention}, etc. In this section, we divide the existing learning-based algorithms into three categories: static, dynamic, and semi-dynamic~\cite{NNLCB}.

\subsubsection{\textbf{Static Methods}}
The static LGC first inputs the whole file as training data to the prediction model, updates the parameters by constant iteration to reduce the difference between the predicted and real symbols, and saves the model parameters at the end of training. In the data compression phase, compressor load and frozen the model parameters, output probability distribution and combine it with an entropy encoder to achieve lossless compression. DeepZip~\cite{DeepZip} and DecMac~\cite{liu2019decmac} use RNN and LSTM as the predictive models respectively. The number of pre-training epochs can be set manually to balance the running time and compression ratio. With the rapid development of large language models, compressors such as LLMZip~\cite{LLMZip} and LMIC~\cite{LMIC} that use large models as predictors have also emerged. Although sufficient prior knowledge allows them to compress various types of data very well, the large models also lead to their extremely slow compression speed.

\subsubsection{\textbf{Dynamic Methods}}
Dynamic LGCs do not need to pre-train on the to-be-compressed LRGS data. They update model parameters in real time during the compression process to minimize the difference between the predicted and target symbol. CMIX~\cite{CMIX} mixes the probability distribution of more than 2,000 models, including dedicated models for different data types, which results in the best compression ratio, but the very low throughput. NNCP~\cite{NNCP} and lstm-compress~\cite{lstm-compress} use Long-short Term Memory (LSTM)~\cite{LSTM} as a probability predictor, where NNCP pre-processes the data to extract repetitive patterns, which improves the compression ratio while increasing the time overhead. TRACE~\cite{TRACE} uses Performer~\cite{choromanski2020rethinking} as a predictor and uses byte-grouping to reduce the length of context to improve the throughput. OREO~\cite{OREO} and PAC~\cite{PAC} use multilayer perception combined with ordered mask. MSDLC~\cite{MSDLC} applies the state-of-the-art model xLSTM~\cite{beck2024xlstm} and designs an extended mapping coding algorithm for pre-processing data in order to reduce the data length and increase the throughput. MSDZip~\cite{MSDZip} introduces Deep Mixing Block and Spatial Gating Unit~\cite{liu2021pay} on the basis of PAC to further improve the compression ratio, and also adopts the designed stepwise-parallel strategy to greatly improve the throughput.

\subsubsection{\textbf{Semi-Dynamic Methods}}
Semi-dynamic based algorithms combine the features of both static and dynamic algorithms. DZip~\cite{DZip} is a typical semi-dynamic compressor, which first pre-trains the data using a static Bootstrap model and saves the model parameters. Then it loaded the model parameters to turn on compression, and applies a dynamic Support model to realize the update of parameters to further improve the compression ratio.

\subsection{\textbf{Traditional LRGS-Dedicated Compressors}}
Enano~\cite{enano} is a typical Tradition Dedicated Compressor (TDC), it parses the input FastQ file in the form of blocks, where the boundaries of each block are dynamically determined to ensure that each read is fully contained within a single block. Each block has three streams: read identifiers, base call sequences, and quality score sequences. For each stream, a specific context model is used to determine the probability distribution of each symbol, which is then passed to an arithmetic encoder~\cite{witten1987arithmetic} to generate the output bit-stream. These probability distributions are dynamically updated as the file is sequentially compressed. 

NanoSpring~\cite{meng2023reference} is a novel lossless reference-free compressor explicitly designed for nanopore sequencing long-read data. NanoSpring employs an approximate assembly approach and demonstrates improved compression performance on higher-quality datasets. Compared to general-purpose compressors and Enano~\cite{enano}, NanoSpring achieves a notable 3-6 fold improvement in compression ratio on the human whole-genome dataset. Furthermore, NanoSpring exhibits competitive compression ratios and resource utilization compared to the state-of-the-art tool CoLoRd~\cite{kokot2022colord} while significantly outperforming it regarding decompression speed.

CoLoRd~\cite{kokot2022colord} has received substantial subsequent support and maintenance. It is designed to effectively compress ONT and PacBio long-reads. CoLoRd is based on overlap graphs~\cite{myers2005fragment} commonly utilized by long-read assemblers~\cite{li2016minimap,koren2017canu}. However, the objective of the CoLoRd algorithm is not to identify actual overlaps but rather to identify overlaps that facilitate efficient compression of read differences. This strategy allows for a trade-off in accuracy to enhance the performance in terms of compression throughput. Additionally, the CoLoRd algorithm supports complete FastQ file compression and parallel acceleration on multi-core CPUs.

FastqCLS~\cite{fastqcls} supports comprehensive compression of FastQ files while facilitating multi-threaded parallel processing. FastqCLS offers lossless reference-free compression for sequences of varying lengths, including long and short sequences. The fundamental principle of the FastqCLS algorithm revolves around a sequence rearrangement computation model developed by the authors. By reordering the sequencing reads and employing the ZPAQ~\cite{ZPAQ} text compression algorithm with contextual modelling, FastqCLS achieves significant compression gains.

Genozip~\cite{genozip} is a comprehensive and versatile compression program designed specifically for genetic information. Genozip is a software designed for genomic compression, providing universality (support for all standard genomic file formats), high compression ratios, speed, feature richness and extensibility. Genozip delivers high-performance compression for widely used genomic data formats in genomics research, namely FastQ, SAM/BAM/CRAM~\cite{li2009sequence}, VCF~\cite{danecek2011variant}, GVF~\cite{reese2010standard}, FastA, PHYLIP and 23andMe formats. 

Spring~\cite{spring} utilizes a hash-based sequence reordering algorithm first used in the HARC~\cite{harc} algorithm. However, the Spring compressor allows for variable-length compression of long-read data, which HARC does not support. 

{
\small
\begin{table*}[!]
\renewcommand\arraystretch{1}
\centering
\caption{Detailed information of all used datasets}
\begin{tabular}{llllrrrrr}
\hline
\textbf{ID} & \textbf{Organism} & \textbf{Name} & \textbf{Platform} &\textbf{FileSize (B)} & \textbf{BaseNum} & \textbf{ReadNum} & \textbf{MaxLen} & \textbf{AvgLen}  \\ \hline
D1 & Human Lung & ERR3077524 & NanoPore (MinION) & 25579723 & 24200685 & 23968 & 4699 & 1010 \\
D2 & SARS-CoV-2 & ERR11274574     & NanoPore (MinION)     & 38575998      & 36605898     & 34758         & 1499        & 1053        \\
D3 & Sediment   & SRR25685106     & NanoPore (MinION)     & 46546159      & 45029637     & 25892         & 11430       & 1739        \\
D4 & Human Lung & ERR3077535      & NanoPore (MinION)     & 61025715      & 59655667     & 23813         & 10399       & 2505        \\
D5 & Alphaherpesvirus       & ERR2708436      & NanoPore (MinION)     & 148511039     & 142000000    & 114447        & 30301       & 1242        \\
D6 & Alphaherpesvirus       & ERR2708427      & NanoPore (MinION)     & 278861604     & 266000000    & 218903        & 10276       & 1217        \\
D7 & Vibrio Cholerae        & ERR13351229     & NanoPore (GridION)    & 775252173     & 769000000    & 197733        & 68098       & 3890        \\
D8 & Escherichia Coli       & SRR25689478     & NanoPore (GridION)    & 2918639845    & 2890000000   & 501216        & 273767      & 5763        \\
D9 & Human Metagenome       & DRR317940       & NanoPore (PromethION) & 615332188     & 610000000    & 195313        & 65111       & 3124        \\
D10 & Homo Sapiens & DRR454989       & NanoPore (PromethION) & 4424406665    & 4200000000   & 3859862       & 168289      & 1088        \\
D11 & Drosophila & SRR1204468      & PacBio (RS)           & 1007829024    & 1000000000   & 166625        & 57717       & 6026        \\
D12 & Yeast      & yeast\_filtered & PacBio (RS)           & 1420763966    & 1400000000   & 232230        & 30164       & 6040        \\
D13 & 48 Plex    & bc1019          & PacBio (Sequel II)    & 3287517449    & 3270000000   & 355254        & 112631      & 9210        \\
D14 & 48 Plex    & bc1099          & PacBio (Sequel II)    & 5201666609    & 5180000000   & 572972        & 95353       & 9035        \\
D15 & Porphyrio Hochstetteri & SRR25655962     & PacBio (HiFi)         & 6021004601    & 6000000000   & 434182        & 45123       & 13809       \\
D16 & Lemur Catta            & SRR25750949     & PacBio (HiFi)         & 8736567109    & 8710000000   & 538247        & 44756       & 16173  \\ \hline
\end{tabular}
\label{tab:dataset}
\begin{flushleft}
\textbf{NOTE}: \textbf{BaseNum} and \textbf{ReadNum} represent the numbers of the Bases and Reads in dataset, respectively; \textbf{MaxLen} and \textbf{AvgLen} denote the maximum and average read length, respectively.
\end{flushleft}
\end{table*}
}

The GeCo series comprises GeCo~\cite{GeCo}, GeCo2~\cite{GeCo2}, and GeCo3~\cite{GeCo3}, proposed by Diogo Pratas et al. in 2016, 2019, and 2020, respectively. The GeCo algorithm series is primarily designed for compressing DNA string sequences. In our benchmark tests, we preserved a 32-bit sequence length index for the GeCo algorithm series to ensure lossless recovery of the original data. Within the GeCo algorithm series, GeCo introduces novel Extended Finite-Context Models (XFCMs) that allow tolerance for substitution errors, enabling reference-free and reference-based DNA sequence compression. GeCo2 is an improved version of GeCo, where the authors enhance the model blending by assigning a specific decay factor to each context model or tolerant context model. Furthermore, GeCo2 incorporates specific cache hash sizes and can exclusively run context models with reverse repetitions. Building upon the GeCo and GeCo2 algorithms, the research team developed GeCo3, a novel genome sequence compressor that utilizes neural networks to blend multiple contexts and substitution-tolerant context models.

The Nucleotide Archival Format (NAF)~\cite{NAF} operates by dividing the input FastQ file into multiple data streams and processing each stream separately. In NAF, sequences are concatenated, and their lengths are stored separately. Then, the combined sequence is transformed into a 4-bit encoding, with each byte storing two nucleotides. Finally, the generated data streams are compressed using the ZSTD~\cite{zstd} general-purpose compressor.

MFCompress~\cite{MFCompress} was specifically designed for compressing FastA and multi-FastA files. Compared to GZip and other tools used for multi-FastA files, MFCompress can provide an additional average compression gain of nearly 50\%, effectively doubling the available storage space. To compress data, we simply convert the reads file into a FastA file and record the sequence lengths. MFCompress uses single finite-context models for encoding the header text, as well as multiple competing finite-context models for encoding the DNA sequences~\cite{MFCompress,green}.

\subsection{\textbf{Learning-based LRGS-Dedicated Compressors}}
Similar to LGCs, learning-based Dedicated Compressors (LDCs) also include two core components, that is, prediction model and entropy encoder. DeepDNA~\cite{wang2018deepdna} is a genomic data compression algorithm based on hybrid convolutional and RNN. It uses convolutional layers to capture local features of the genome and RNN to capture long-term dependencies in order to accurately predict the probability distribution of target bases.
DNA-BiLSTM~\cite{cui2020compressing} further improves the compression ratio by replacing the RNN with an attention-based bi-directional LSTM model based on DeepDNA.
CompressBERT~\cite{yang2023rethinking} uses BERT~\cite{devlin2019bert} as a predictive model.
GenCoder~\cite{sheena2024gencoder} applies a convolutional auto-encoder to achieve reference-free compression of genome sequences.
GeneFormer~\cite{cui2024geneformer} introduces an improved Transformer encoder with latent array to remove dependencies between nucleotide sequences, and applies a multi-level grouping approach to accelerate compression.
LEC-Codec~\cite{sun2024lec} uses Group of Bases and LSTM to achieve compression.
AGDLC~\cite{AGDLC} uses a multiple (\textit{s},\textit{k})-mer mixer for extracting redundant information from genomic data in multiple dimensions, and introduces xLSTM~\cite{beck2024xlstm} as the prediction model.
DeepGeCo~\cite{DeepGeCo} takes dynamic transformer as backbone, on top of it, introducing BiLSTM-based static model to solve cold-start problem for enhancing compression ratio.

\section{Benchmark Evaluation}~\label{S3}
All experiments were conducted on a Linux server (Ubuntu 20.04.6 LTS) equipped with 4 $\times$ Intel Xeon Silver 4310 CPUs (2.10 GHz, 48 cores in total), 4 $\times$ NVIDIA GeForce RTX 4090 GPUs (16,384 CUDA cores, 24 GB of GPU memory), and 256 GB of DDR4 RAM.
{
\small
\begin{table*}[]
\renewcommand\arraystretch{0.9}
\centering
\caption{Detailed information of all tested compressors}
\begin{tabular}{lcccccccc}
\hline
\textbf{Compressor} & \textbf{Method Type} & \textbf{Year} & \textbf{Used Version} & \textbf{Language} & \textbf{Technologies} & \textbf{Multi-threads?} & \textbf{S/L/A} & \textbf{Hyper-parameters} \\ \hline
Brieflz~\cite{Brieflz} & TGC & 2020          & 1.3.0     & C/C++ & LZ77 & NO          & Y/Y/Y          & -9           \\
Brotli~\cite{Brotli}   & TGC & 2018          & 1.1.0     & C/C++ & LZ77, HC, CM     & NO          & Y/Y/Y          &  \\
BSC~\cite{BSC}         & TGC & 2016          & 3.3.4     & C/C++ & BWT  & NO          & Y/Y/Y          & -e2          \\
BZip2~\cite{Bzip2}     & TGC & 2010          & 1.0.6     & C/C++ & BWT, MTF     & NO          & Y/Y/Y          & -9           \\
GZip~\cite{Gzip}       & TGC & 1996          & 1.10      & C/C++ & LZ77, HC     & NO          & Y/Y/Y          &  \\
Lizard~\cite{Lizard}   & TGC & 2017          & 1.0       & C/C++ & LZ4, HC          & NO          & Y/Y/Y          & -49          \\
LZ4~\cite{LZ4}         & TGC & 2019          & 1.10.0    & C/C++ & LZ77 & YES         & Y/Y/Y          & -12 -BD      \\
LZMA~\cite{LZMA}       & TGC & 2022          & 24.08     & C/C++ & LZ77, AC         & NO         & Y/Y/Y          & -mx9         \\
LZMA2~\cite{LZMA}      & TGC & 2022          & 24.08     & C/C++ & LZ77, AC         & YES         & Y/Y/Y          & -mx9         \\
LZO~\cite{LZO}         & TGC & 1996          & 2.10      & C/C++ &    LZ77  &    NO       & Y/Y/Y          & -9           \\
LZOP~\cite{lzop}       & TGC & 1997          & 1.04      & C/C++ & LZ77 & NO          & Y/Y/Y          & -9           \\
PBZip2~\cite{PBZIP2}   & TGC & 2015          & 1.1.13    & C/C++ & BWT, HC          & YES         & Y/Y/Y          & -9 -m2000    \\
PigZ~\cite{pigz}       & TGC & 2013          & 2.8       & C/C++ & LZ77, Zopfli     & YES         & Y/Y/Y          & -9           \\
PPMD~\cite{PPMD}       & TGC & 2022          & 24.08     & C/C++ & PPM, AC          & NO         & Y/Y/Y          & -mx9         \\
SnZip~\cite{Snzip}     & TGC & 2016  & 1.0.5     & C/C++     & LZ77  & NO       & Y/Y/Y          &  \\
ZPAQ~\cite{ZPAQ}       & TGC & 2009          & 7.15      & C/C++ & CM, AC           & YES         & Y/Y/Y          & -m5          \\
ZSTD~\cite{zstd}       & TGC & 2021          & 1.5.6     & C/C++ & LZ77, HC         & NO          & Y/Y/Y          & -19          \\
CMIX~\cite{CMIX}       & LGC & 2014          & 21        & C/C++ & DL, CM, AC       & NO          & Y/Y/Y          &  \\
NNCP~\cite{NNCP}       & LGC & 2019          & 24.6.5    & C/C++ & DL, AC           & NO          & Y/Y/Y          &  \\
lstm-compress~\cite{lstm-compress} & LGC & 2016          & 3         & C/C++ & DL, AC           & NO          & Y/Y/Y          &   \\
DeepZip~\cite{DeepZip} & LGC & 2019          & 1.0       & Python            & DL, AC           & NO          & Y/Y/Y          &  -bs 512 -ts 64\\
DZip~\cite{DZip}       & LGC & 2021          & 1.0       & Python            & DL, AC           & NO          & Y/Y/Y          &  -bs 64 -ts 64\\
TRACE~\cite{TRACE}     & LGC & 2022          & 1.0       & Python            & DL, AC           & NO          & Y/Y/Y          &  -bs 512  \\
PAC~\cite{PAC}         & LGC & 2023          & 1.0       & Python            & DL, AC           & NO          & Y/Y/Y          &  -bs 512  \\
MSDLC~\cite{MSDLC}     & LGC & 2025          & 1.0       & Python            & DL, AC           & NO          & Y/Y/Y          &  -bs 512 -ts 32 \\
MSDZip~\cite{MSDZip}   & LGC & 2025          & 1.0       & Python            & DL, AC           & YES         & Y/Y/Y          &  -bs 512 -ts 16  \\ 
CoLoRd~\cite{kokot2022colord}  & TDC    & 2022          & 1.2.1     & C/C++ & KMC, OG, Zlib    & YES         & N/Y/N          &  \\
Enano~\cite{enano}     & TDC & 2020          & 1.0       & C/C++ & FCMs, AC         & YES         & N/Y/N          & \\
GeCo~\cite{GeCo}       & TDC & 2016          & 2.1       & C/C++ & FCMs, XCFMs      & NO          & Y/Y/Y          & -l 5         \\
GeCo2~\cite{GeCo2}     & TDC & 2019          & 1.1       & C/C++ & FCMs, XCFMs      & NO          & Y/Y/Y          & -l 5         \\
MFCompress~\cite{MFCompress}  & TDC      & 2014          & 1.01      & C/C++ & FCMs, AC         & NO          & Y/Y/Y          &  \\
NAF~\cite{NAF}         & TDC & 2019          & 1.3.0     & C/C++ & ZSTD & NO          & Y/Y/Y          &  \\
Spring~\cite{spring}   & TDC      & 2019          & 1.1.1      & C/C++ & RO, BSC, LZMA2         & YES          & Y/Y/N          & -l  \\
AGDLC~\cite{AGDLC}     & LDC & 2025          & 1.0       & Python            & DL, AC           & NO          & Y/Y/Y          & \\ 
GeCo3~\cite{GeCo3}     & LDC & 2020          & 1.0       & C/C++ & FCMs, XCFMs, NN  & NO          & Y/Y/Y          & -l 5         \\ \hline
\end{tabular}
\label{tab:compressor}
\begin{flushleft}
\textbf{NOTE}: \textbf{TGC}, \textbf{LGC}, \textbf{TDC}, and \textbf{LDC} represent traditional general-purpose, learning-based general-purpose, traditional dedicated, and learning-based dedicated compressors, respectively. \textbf{PPM} (Prediction by Partial Match): order-n, modeled in longest context matched, but dropping to lower orders for byte counts of 0; \textbf{Zopfli}: a very highly optimized and slow implementation of deflate; \textbf{HC}: Huffman Coding; \textbf{AC}: Arithmetic Coding; \textbf{FCMs} (Finite Context Models); XCFMs (Extended Finite Context Models); \textbf{NN}: Neural Networks; \textbf{RO} (ReOrder): reorder DNA reads according to sequence similarity; \textbf{BitS}: Bits Stream; \textbf{DNS}: De Novo Assembly; \textbf{KMC}: K-Mer Counting; \textbf{OG} (Overlap Graphs): overlap graphs often employed by long-read assemblers; \textbf{DL}: Deep Learning; \textbf{S/L/A} (Short/Long/Assembly Sequences): whether corresponding compression patterns are supported; \textbf{MTF}: Move-To-Front.
\end{flushleft}
\end{table*}
}
\subsection{Setup}
\subsubsection{\textbf{Experimental Datasets}}
We collected 16 real sequencing datasets of different types from two mainstream sequencing platforms, ONT and PacBio on European Bioinformatics Institute Website~\cite{EBI2023}. We selected three types of data including MinION (D1$\sim$D6), GridION(D7 and D8), and PromethION (D9 and D10) from the ONT, and three types of data including PacBio RS (D11 and D12), Sequel II (D13 and D14), and HiFi (D15 and D16) from the PacBio. 
The detailed information of all datasets is shown in Table~\ref{tab:dataset}.
\subsubsection{\textbf{Benchmark Compressors}}
We reproduce and test 35 LRGS data compressors, including 17 TGC-based, 9 LGC-based, 7 TDC-based, and 2 LDC-based methods. 
Detailed information for all benchmark compressors is shown in Table~\ref{tab:compressor}.

\subsubsection{\textbf{Evaluation Metrics}}
We use Compression Ratio (\textit{CR}) and Storage Saving Percentage (\textit{SSP}) to evaluate the compression effectiveness of the compressor~\cite{sun2024pqsdc,NNLCB,sun2023sr2c,sun2023pmffrc}. The \textit{CR} and \textit{SSP} are defined as Eq.~(\ref{eq:CR}) and Eq.~(\ref{eq:SSP}), respectively.

\begin{equation}
\label{eq:CR}
    CR =\frac{Size_{compressed}}{Size_{source}}\times 8 \; bits/base
\end{equation}

\begin{equation}
\label{eq:SSP}
    SSP=(1-\frac{Size_{compressed}}{Size_{source}})\times 100 \%
\end{equation}
where $Size_{source}$ and $Size_{compressed}$ represent the file size before and after compression, respectively.

{
\small
\begin{table*}[]
\renewcommand\arraystretch{0.8}
\centering
\caption{Benchmark results of all compressors on tested real-world datasets}
\begin{tabular}{llllllllll}
\toprule
\textbf{Tested Algorithm} & \textbf{\textit{AvgCR \textdownarrow}} & \textbf{\textit{WavgCR \textdownarrow}} & \textbf{\textit{AvgSSP \textuparrow}} & \textbf{\textit{WavgSSP \textuparrow}} & \textbf{\textit{RP \textdownarrow}} & \textbf{\textit{TotalCT \textdownarrow}} & \textbf{\textit{TotalDT \textdownarrow}} & \textbf{\textit{MaxCPM \textdownarrow}} & \textbf{\textit{MaxDPM \textdownarrow}}  \\
& (bits/base)    & (bits/base)     & (\%)            & (\%)             & (\%)        & (Hour)            & (Hour)            & (KB)           & (KB) \\ \hline
\multicolumn{10}{l}{\textbf{Results on all 16 datasets}}                                                                                              \\
CoLoRd     & \textbf{1.050} $^{\#1}$ & \textbf{0.902} $^{\#1}$ & \textbf{86.887} $^{\#1}$ & \textbf{88.728} $^{\#1}$  & 49.235 & 4.026   & 0.264   & 8583768  & 3943400  \\
GeCo       & \textbf{1.189} $^{\#2}$ & \textbf{1.326} $^{\#2}$ & \textbf{85.149} $^{\#2}$ & \textbf{83.450}  $^{\#2}$  & 44.765 & 9.647   & 9.755   & 5049816  & 5049520  \\
GeCo3      & \textbf{1.281} $^{\#3}$   & 1.542 & \textbf{83.994} $^{\#3}$ & 80.719  & 40.961 & 18.946  & 16.471  & 527916   & 527604   \\
MFCompress & 1.344       & \textbf{1.468} $^{\#3}$       & 83.197 & \textbf{81.651} $^{\#3}$  & 39.271 & 2.312   & 2.523   & 2507756  & 2386176  \\
GeCo2      & 1.373       & 1.613       & 82.842 & 79.833  & 37.352 & 3.193   & 3.348   & 527900   & 527576   \\
NAF        & 1.403       & 1.613       & 82.460 & 79.841  & 39.181 & 6.820   & \textbf{0.020} $^{\#3}$   & 3694580  & 439080   \\ 
Brotli     & 1.425       & 1.609       & 82.182 & 79.893  & 34.724 & 32.755  & 0.070   & 272504   & 19460    \\
LZMA       & 1.445       & 1.648       & 81.934 & 79.374  & 37.989 & 12.713  & 0.083   & 692904   & 70892    \\
LZMA2      & 1.445       & 1.648       & 81.932 & 79.373  & 37.995 & 12.427  & 0.082   & 693064   & 71000    \\
ZPAQ       & 1.551       & 1.687       & 80.614 & 78.916  & 28.398 & 23.800  & 23.982  & 896808   & 811280   \\
Spring     & 1.561       & 1.711       & 80.505 & 78.640  & 28.563 & 1.196   & 0.719   & 1009824  & 969688   \\
BSC        & 1.576       & 1.734       & 80.312 & 78.330  & 27.672 & \textbf{0.060}$^{\#1}$   & 0.043   & 14275248 & 14346848 \\
ZSTD       & 1.637       & 1.788       & 79.543 & 77.656  & 28.799 & 8.047   & \textbf{0.016} $^{\#2}$   & 232116   & 13396    \\
PPMD       & 1.698       & 1.846       & 78.780 & 76.922  & 27.165 & 2.349   & 2.425   & 268556   & 267620   \\
Enano      & 1.719       & 1.819       & 78.513 & 77.260  & \textbf{19.008} $^{\#2}$ & \textbf{0.572} $^{\#3}$   & 0.791   & 46260    & 42008    \\
Lizard     & 1.773       & 1.892       & 77.834 & 76.356  & 25.215 & 6.648   & 0.021   & 175100   & 39832    \\
BZip2      & 1.890       & 2.014       & 76.374 & 74.822  & 23.415 & 0.898   & 0.447   & 8100     & 5024     \\
PBZip2     & 1.892       & 2.016       & 76.345 & 74.806  & 23.285 & 0.928   & 0.447   & 10908    & 7308     \\
GZip       & 2.024       & 2.082       & 74.714 & 73.991  & \textbf{21.901} $^{\#3}$ & 11.351  & 0.064   & \textbf{2148} $^{\#1}$     & \textbf{1968} $^{\#2}$     \\
PigZ       & 2.025       & 2.084       & 74.682 & 73.952  & 21.942 & 7.561   & 0.050   & 2672     & 2284     \\
Brieflz    & 2.146       & 2.244       & 73.192 & 71.954  & 23.815 & 3.968   & 0.122   & 24052    & 3040     \\
LZ4        & 2.497       & 2.590       & 68.788 & 67.624  & 22.555 & 7.943   & \textbf{0.011} $^{\#1}$   & 13056    & 36348    \\
LZO        & 2.578       & 2.686       & 67.800 & 66.460  & 22.511 & 17.593  & 0.033   & \textbf{2372} $^{\#2}$     & \textbf{1736} $^{\#1}$     \\
LZOP       & 2.578       & 2.685       & 67.801 & 66.460  & 22.522 & 18.923  & 0.033   & \textbf{2624} $^{\#3}$     & \textbf{2152} $^{\#3}$     \\
SnZip      & 3.609       & 3.655       & 54.883 & 54.305  & \textbf{9.832} $^{\#1}$  & \textbf{0.085}$^{\#2}$   & 0.056   & 3424     & 3404     \\
\multicolumn{10}{l}{\textbf{Results on 6 MinION datasets (due to high running time)}} \\ 
CMIX                               & 1.299       & 1.504       & 83.775 & 81.220  & 45.161 & 218.861 & 221.631 & 24186092 & 24186028 \\
NNCP                               & 1.363       & 1.537       & 82.965 & 80.784  & 42.333 & 19.655  & 22.112  & 302308   & 302244   \\
MSDZip                             & 1.415       & 1.588       & 82.313 & 80.160  & 38.245 & 3.710   & 5.269   & 5003868  & 5874872 \\ 
PAC                                & 1.438       & 1.595       & 82.045 & 80.084  & 36.228 & 3.371   & 5.724   & 5005940  & 5732072  \\
MSDLC                              & 1.473       & 1.605       & 81.607 & 79.964  & 32.234 & 5.714   & 6.748   & 23377316 & 6295000  \\ 
TRACE                              & 1.473       & 1.641       & 81.595 & 79.491  & 34.813 & 4.412   & 6.305   & 5006092  & 5781852  \\
DZip                               & 1.625       & 1.633       & 79.690 & 79.584  & 39.366 & 19.891  & 6.710   & 5574340  & 4132124  \\
DeepZip                            & 1.652       & 1.666       & 79.347 & 79.172  & 37.781 & 15.308  & 2.171   & 18903952 & 5159844  \\
AGDLC                              & 1.802       & 1.812       & 77.476 & 77.354  & 3.437  & 5.971   & 7.254   & 11826492 & 3763124  \\
lstm-compress                      & 6.815       & 8.104       & 14.825 & -1.297  & 41.884 & 16.310  & 16.211  & 6716     & 6792     \\
\bottomrule
\end{tabular}
\label{tab:result}
\begin{flushleft}
\textbf{NOTE}: The three best results are in bold, with the symbol "\#" indicating the rankings. \textbf{\textit{TotalCT}} and \textbf{\textit{TotalDT}} denote the total compression and decompression running time, respectively. \textbf{\textit{MaxCPM}} and \textbf{\textit{MaxDPM}} represent the maximum compression and decompression peak memory usage, respectively. 
\end{flushleft}
\end{table*}
}

In addition, in order to comprehensively evaluate the overall compression effect on different datasets, we also propose the Average or Weighted Compression Ratio (\textit{Avg/WavgCR}), Average or Weighted Storage Saving Percentage (\textit{Avg/WavgSSP}), and Robust Performance (\textit{RP}). The \textit{Avg/WavgCR}, \textit{Avg/WavgSSP}, and \textit{RP} are defined as follows~\cite{sun2024lrcb,NNLCB,xing2017gtz,sun2024pqsdc}:

\begin{equation}
\label{eq:Avg/WavgCR}
    Avg/WavgCR=\sum_{i=0}^{N-1}\alpha_i\times CR_i
\end{equation}

\begin{equation}
\label{eq:Avg/WavgSSP}
    Avg/WavgSSP=\sum_{i=0}^{N-1}\alpha_i\times SSP_i
\end{equation}

\begin{equation}
\label{eq:RP}
    RP=\frac{\sqrt{\sum_{i=0}^{N-1}(CR_i-AvgCR)^2}}{\sqrt{N}\times AvgCR}\times 100\%
\end{equation}
where $CR_i$ and $SSP_i$ denote the \textit{CR} and \textit{SSP} of the compressor on the \textit{i}-th dataset, respectively. The variable $N$ denotes the number of benchmark datasets. The variable $\alpha_i$ denotes the weighting factor of the \textit{i}-th dataset, for \textit{AvgCR} and \textit{AvgSSP}, we set $\alpha_i=\frac{1}{N}$; and for the \textit{WavgCR} and \textit{WavgSSP}, we set $\alpha_i=\frac{Size_{compressed}^i}{\sum_{j=0}^{N-1}Size_{compressed}^j}$. \textit{RP} is calculated in the same way as the Coefficient of Variation (CV)~\cite{xing2017gtz}, which is calculated as the ratio of standard deviation and mean value multiplied by 100\%. The smaller the \textit{RP} value, the less the algorithm is disturbed by the probability distribution of the datasets and the higher the compression robustness performance. Besides, we evaluate the efficiency of the algorithms in terms of compression and decompression time overhead and peak memory usage.

\subsection{Result Analysis}
Table~\ref{tab:result} shows the test results of all the algorithms. Due to the large time overhead of current learning-based compressors, it is difficult to complete the test on all datasets. Therefore we only tested the learning-based algorithms on ONT MinION datasets.

\subsubsection{\textbf{Compression Ratio \& Robust Performance}}
Table~\ref{tab:result} shows that the overall \textit{CRs} of the LRGS-dedicated compressors are overall better than that of the general-purouse algorithms, especially CoLoRd, which achieves the best compression effects in both \textit{AvgCR} (1.050) and \textit{WavgCR} (0.902). 
The rankings of the two metrics, \textit{AvgCR} and \textit{WavgCR}, are slightly different, which is due to the fact that \textit{WavgCR} takes into account the differences in overall compression results caused by different sizes. For example, GeCo2 (1.613) and NAF (1.613) have slightly worse \textit{WavgCR} than the traditional algorithm Brotli (1.609). SSP is a more intuitive representation of \textit{CR}, and thus \textit{AvgSSP} and \textit{WavgSSP} are ranked the same as \textit{AvgCR} and \textit{WavgCR}, respectively. 

As for \textit{RP}, SnZip has the worst compression ratio (3.609 for \textit{AvgCR} and 3.655 for \textit{WAvgCR}), instead has the best compression robustness (9.832 \%), which is due to the fact that SnZip emphasizes more on the compression speed and thus is not sensitive to the dataset, leading to its greater robustness. 
In contrast, the dedicated compressor with the best \textit{CR} possesses the largest \textit{RP} value, which means that specialized compression algorithms aiming to maximize the use of redundant features are more sensitive to the probability distribution of nucleotide bases of the compressed data. 
Since dedicated compressors require additional computations to utilize redundant features in sequencing data to improve the compression ratio, the time overhead and peak memory overhead of dedicated compressors are usually higher than that of general-purpose algorithms. 

\subsubsection{\textbf{Time \& Memory Consumption}}
As shown in Table~\ref{tab:result}, the total compression\&decompression time of CoLoRd (4.291 Hours), Spring (1.915 Hours), Enano (1.363 Hours), GeCo2 (6.541 Hours), MFCompress (4.835 Hours), and NAF (6.840 Hours) is much less than that of some traditional algorithms such as Brotli (32.825 Hours), ZPAQ (47.782 Hours), when they are all single-threaded. BSC is closer to SnZip in terms of runtime, but the \textit{CR} of BSC (1.576 for \textit{AvgCR} and 1.734 for \textit{WAvgCR}) is significantly higher than that of SnZip (3.609 for \textit{AvgCR} and 3.655 for \textit{WAvgCR}), while the peak memory overhead of BSC (13.614 GB for compression and 13.682 GB for decompression) is the highest among all algorithms. 
Dedicated compressors usually have higher peak memory overheads, for example CoLoRd, which has the best \textit{CR}, has a peak memory overhead of 8.186 GB. GZip (2,148 KBs), LZO (2,372 KBs) and LZOP (2,624 KBs) have the smallest peak memory overheads.

\subsubsection{\textbf{Learning-based Methods on MinION datasets}}
As for the test results of the learning-based algorithms on the 6 MinION datasets, we can see that CMIX~\cite{CMIX} has the best \textit{CR} (1.299 for \textit{AvgCR} and 1.504 for \textit{WavgCR}) due to the fact that it mixes the probability distributions of more than 2,000 models, which includes specialized models for different data types. But at the same time it is much slower than any algorithm, with a total compression and decompression time of more than 400 hours (218.861 hours for compression and 221.631 hours for decompression). lstm-compress~\cite{lstm-compress} has a \textit{WavgCR} value greater than 8, which is an anomaly, i.e., the compressed file is larger than the pre-compressed file. The possible reason is that lstm-compress has the problem of gradient disappearance or explosion of the model during the compression process. DZip~\cite{DZip} introduces a model dynamic update mechanism based on DeepZip~\cite{DeepZip}, and thus obtains a better \textit{CR}, but at the same time has a longer compression time and higher memory usage. AGDLC~\cite{AGDLC} is a dedicated compressor, but it seems to be not very good at compressing ONT data, resulting in unsatisfactory compression, resulting in unsatisfactory compression results. But AGDLC has the best compression robustness (3.437\%) due to its dataset insensitivity. MSDLC~\cite{MSDLC} has the second best \textit{RP} (38.245\%), but has better \textit{CR} than several learning-based algorithms, including lstm-compress, DeepZip, DZip, and AGDLC. MSDZip's \textit{CR} (1.802 for \textit{AvgCR} and 1.812 for \textit{WavgCR}) is second only to CMIX (1.299 for \textit{AvgCR} and 1.504 for \textit{WavgCR}) and NNCP (1.363 for \textit{AvgCR} and 1.537 for \textit{WavgCR}) among LGCs, but the time overhead is 2 and 1 order of magnitude less than CMIX and NNCP, respectively.

\begin{figure}[]
    \centering
    \includegraphics[width=\linewidth]{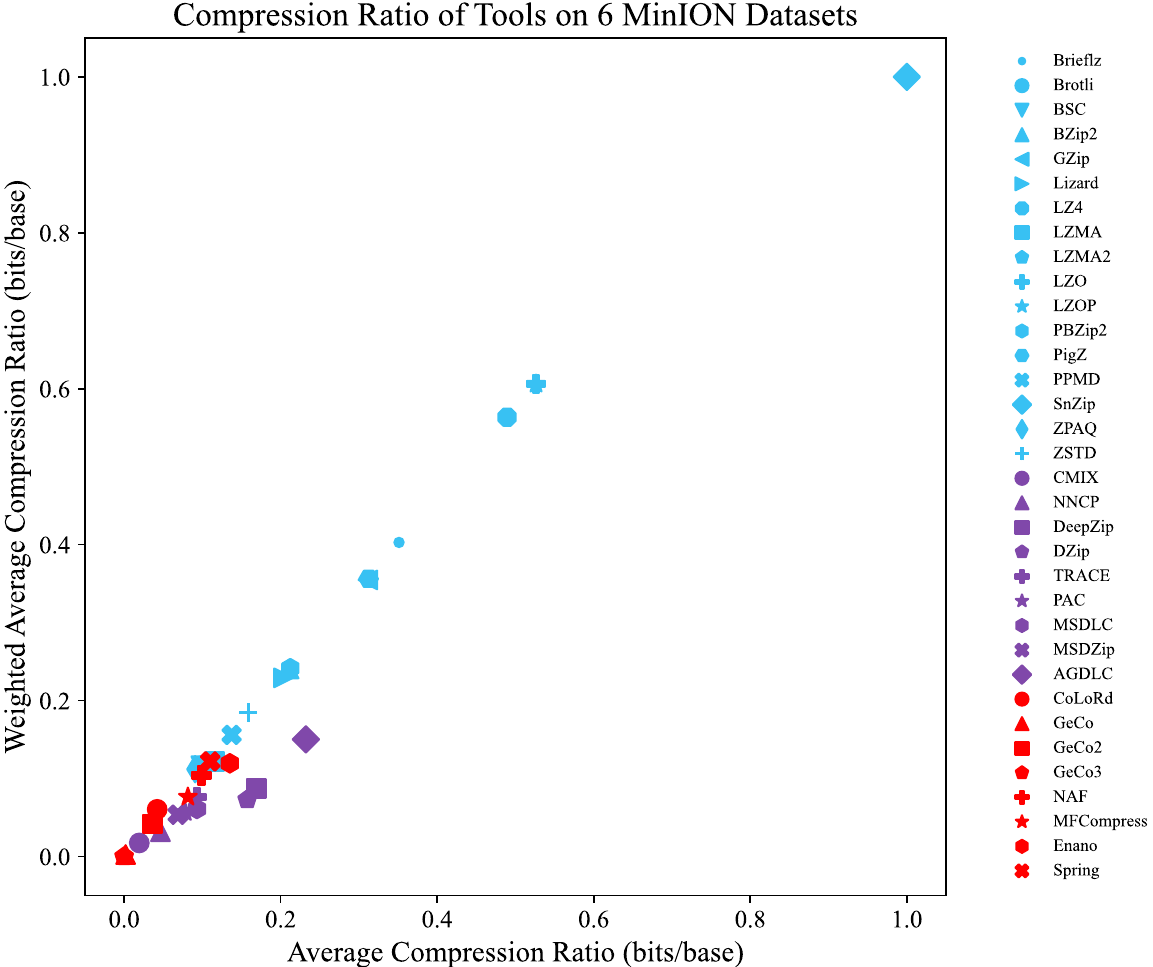}
    \caption{Distribution of compression ratios}
    \label{fig:cr}
\end{figure}

In order to compare all algorithms fairly, we plotted the \textit{AvgCR} and \textit{WavgCR} distributions on the first 6 MinION datasets, as shown in Fig~\ref{fig:cr}. We set the markers for traditional, learning-based and dedicated algorithms to three colors. The closer to the origin means the algorithm has better compression effect. To avoid the effect of outliers on the distribution, we removed the result of lstm-compress.

From the Fig.~\ref{fig:cr}, it can be seen that the \textit{CRs} of the dedicated compressors are better than that of learning-based and traditional algorithms on the 6 MinION datasets, especially GeCo and GeCo3, which obtain the best compression ratio. The \textit{CRs} of the learning-based generalized compressors are better than the traditional algorithms, which is attributed to the fact that the deep learning model can better learn more complex patterns in the data to reduce data redundant information to achieve efficient compression.

\begin{figure*}[t]
    \centering
    \begin{minipage}{0.33\textwidth}
        \centering
        \includegraphics[width=\linewidth]{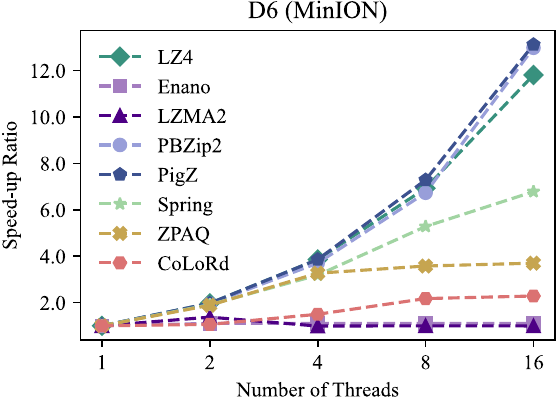}
    \end{minipage}%
    \begin{minipage}{0.33\textwidth}
        \centering
        \includegraphics[width=\linewidth]{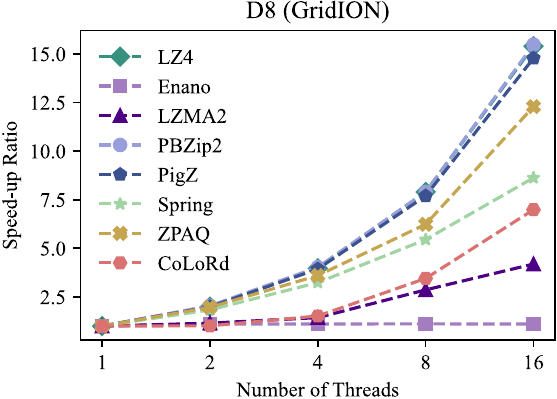}
    \end{minipage}%
    \begin{minipage}{0.33\textwidth}
        \centering
        \includegraphics[width=\linewidth]{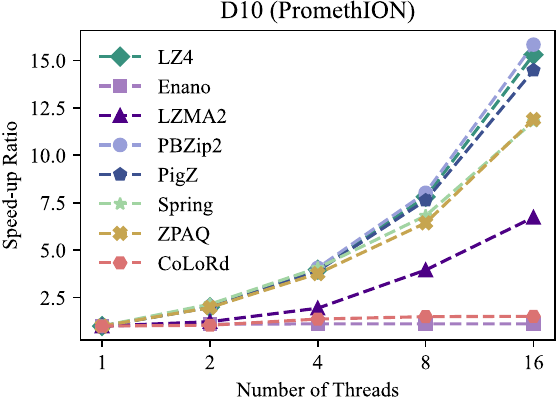}
    \end{minipage}

    \vspace{0.1cm} 

    \begin{minipage}{0.33\textwidth}
        \centering
        \includegraphics[width=\linewidth]{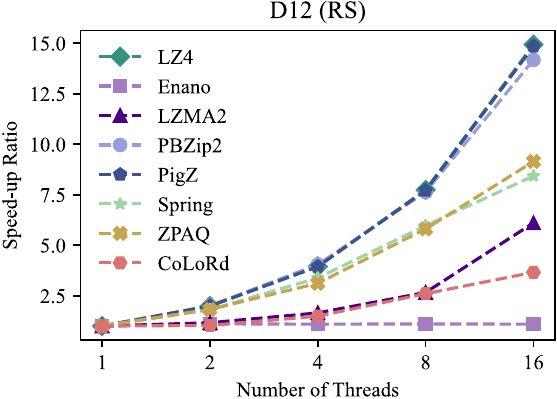}
    \end{minipage}%
    \begin{minipage}{0.33\textwidth}
        \centering
        \includegraphics[width=\linewidth]{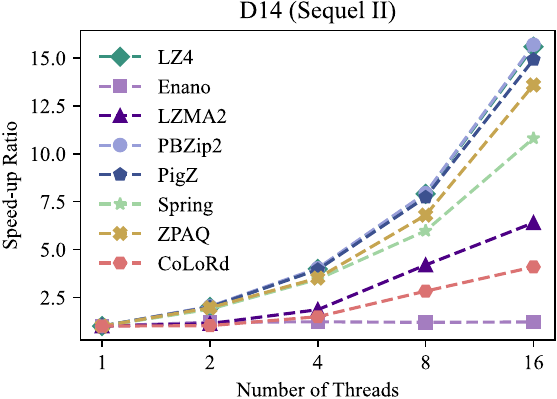}
    \end{minipage}%
    \begin{minipage}{0.33\textwidth}
        \centering
        \includegraphics[width=\linewidth]{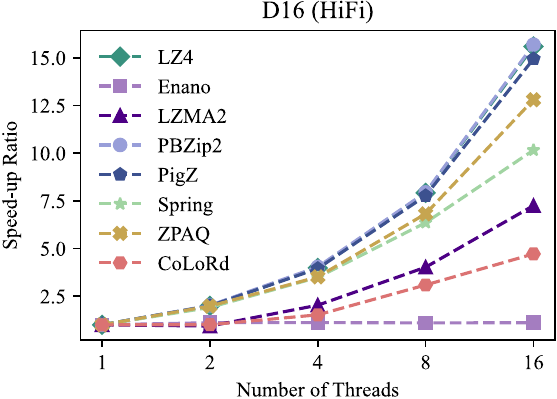}
    \end{minipage}
    \caption{Speedup Ratios of 8 algorithms on 6 MinION datasets}
    \label{fig:sr}
    \vspace{0.2cm}
\end{figure*}

\subsubsection{\textbf{Parallel Acceleration Effect}}
We also tested the parallel effect of the compressors for those algorithms that support multi-threaded parallelism.
Specifically, we tested the speedup ratios of 8 algorithms that support multi-threaded parallel compression  as shown in Fig.~\ref{fig:sr}, including LZ4, LZMA, LZMA2, PBZip2, PigZ, PPMD, ZPAQ, and CoLoRd. The Speedup Ratio (\textit{SR}) is defined as~\cite{cheng2023parallel} :
$SR=\frac{T_{serial}}{T_{parallel}}$,
where $T_{serial}$ and $T_{parallel}$ denote the total compression and decompression time for single and multi-threaded compression respectively. We set the number of threads to 1, 2, 4, 8, and 16. The larger the value of the \textit{SR}, the better the multi-threaded speedup. As shown in Fig.~\ref{fig:sr}, LZ4, PBZip2, and PigZ have a greater upward trend of the folds, indicating that they have the best parallel speedup. ZPAQ and Spring follow them all three, but the speedup of ZPAQ on dataset D6 is not insignificant. LZMA2 has speedups on all datasets except the first one. And CoLoRd has a speedup effect on all datasets except the D10 dataset, where the \textit{SR} is almost the same on the D6 dataset using 16 threads versus 8 off-the-shelf. Enano has a very fast compression and decompression speed. However, its multi-threading acceleration is not very effective.

\section{Future Potential Work}~\label{S4}
Based on the benchmark work in this paper, we summarized the following four valuable recommendations for the future development of long-read big data compressors.

\textbf{1) Parallel Acceleration Compression.}
Hardware acceleration devices, such as FPGA, GPU, and TPU, have shown extensive potential for applications in scientific computing. The design of specialized compression algorithms based on accelerated hardware to address the challenge of compressing large-scale bio-informatics big data holds significant research value~\cite{sun2025pmklc}. 
\textbf{2) Privacy-sensitive Security Compression.}
Unlike ordinary sequencing data, human sample data carries crucial genetic information. The potential leakage of individual privacy information impact healthcare, insurance, and employment~\cite{s1,s2}. The development of secure long-read compression frameworks that integrate bio-data security protection techniques, such as sensitive sequence filtering, secure multiparty computation, and trusted execution environments, hold significant research and practical value. 

\textbf{3) Self-adaptive Parameter Optimization.}
Our investigation revealed that compressors' performance is affected by selected hyper-parameters and data probability distribution. There's limited research on identifying distributions from datasets. Additionally, algorithm configurations are complex and hard for non-computer professionals to understand. Therefore, an automated tuning framework using long-read data's distribution would simplify compressor use and improve performance~\cite{hui2026agentgc}.
\textbf{4) Hardware-algorithm Integration Design.}
Storing extensive bio-sequencing data can be difficult because of the complexity of the application scenarios. For instance, extreme data compression is crucial for long-term backups in large-scale databases. In contrast, real-time data transmission scenarios require a balance between compression, time, and overall performance in terms of memory usage. Therefore, it is valuable to research how to maximize hardware resource utilization in various configurations of compression servers while considering the integrated design of compression algorithms and hardware resources. 

\section{Conclusion}~\label{S5}


We present LRCBench, a benchmark for evaluating reference-free lossless compression of long-read sequencing data. It brings together 16 real-world datasets from diverse sequencing platforms, species, and read lengths, and evaluates 35 compressors using compression ratio, runtime, peak memory, and robustness. The results show that LRGS-dedicated compressors generally achieve stronger compression, while runtime, memory use, and robustness vary across methods and datasets. CoLoRd attains the best overall compression ratios in our tests, whereas lightweight general-purpose tools offer advantages in speed or memory efficiency. We also identify four directions for future work: parallel acceleration, privacy-sensitive compression, self-adaptive parameter optimization, and hardware-algorithm co-design. Together, the benchmark results offer evidence to guide compressor selection and reveal trade-offs that single-metric comparisons can miss. Future updates will expand dataset coverage across platforms and probability distributions and include newly developed compressors, supporting more representative comparisons as long-read technologies evolve.

\newpage
\bibliographystyle{ACM-Reference-Format}
\balance
\bibliography{reference}


\end{document}